\documentclass[twocolumn,prx,floatfix,superscriptaddress]{revtex4}
\usepackage{color}
\usepackage{tabularx}
\usepackage{epsfig}
\usepackage{amsmath}
\usepackage{amssymb}
\usepackage{graphicx}
\usepackage{wasysym}
\usepackage{float}
\usepackage{comment}
\usepackage{paralist}
\usepackage{times}

\begin{document}

\title{Best-case performance of quantum annealers on native spin-glass
benchmarks: \newline How chaos can affect success probabilities}

\author{Zheng Zhu}
\affiliation {Department of Physics and Astronomy, Texas A\&M University,
College Station, Texas 77843-4242, USA}

\author{Andrew J. Ochoa}
\affiliation {Department of Physics and Astronomy, Texas A\&M University,
College Station, Texas 77843-4242, USA}

\author{Stefan Schnabel}
\affiliation{Institut f{\"u}r Theoretische Physik and Centre for
Theoretical Sciences (NTZ), Universit{\"a}t Leipzig, Postfach 100920,
D-04009 Leipzig, Germany}

\author{Firas Hamze}
\affiliation{D-Wave Systems, Inc., 3033 Beta Avenue, Burnaby, British
Columbia, V5G 4M9, Canada}

\author{Helmut G. Katzgraber}
\affiliation{Department of Physics and Astronomy, Texas A\&M University,
College Station, Texas 77843-4242, USA}
\affiliation{Materials Science and Engineering Program, Texas A\&M
University, College Station, Texas 77843, USA}
\affiliation{Santa Fe Institute, 1399 Hyde Park Road, Santa Fe, NM
87501, USA}

\date{\today}

\begin{abstract}

Recent tests performed on the D-Wave Two quantum annealer have revealed
no clear evidence of speedup over conventional silicon-based
technologies.  Here, we present results from classical
parallel-tempering Monte Carlo simulations combined with isoenergetic
cluster moves of the archetypal benchmark problem---an Ising spin
glass---on the native chip topology.  Using realistic uncorrelated noise
models for the D-Wave Two quantum annealer, we study the best-case
resilience, i.e., the probability that the ground-state configuration is
not affected by random fields and random-bond fluctuations found on the
chip.  We thus compute classical upper-bound success probabilities for
different types of disorder used in the benchmarks and predict that an
increase in the number of qubits will require either error correction
schemes or a drastic reduction of the intrinsic noise found in these
devices. We restrict this study to the exact ground state, however, the
approach can be trivially extended to the inclusion of excited states if
the success metric is relaxed.  We outline strategies to develop robust,
as well as hard benchmarks for quantum annealing devices, as well as any
other (black box) computing paradigm affected by noise.

\end{abstract}

\pacs{75.50.Lk, 75.40.Mg, 05.50.+q, 64.60.-i}

\maketitle

\section{Introduction}

Although a useful universal quantum computer
\cite{nielsen:00,nishimori:01} is far from reality at the moment, the
advent of quantum annealing (QA) machines based on quantum adiabatic
optimization techniques
\cite{finnila:94,kadowaki:98,brooke:99,farhi:00,roland:02,santoro:02,das:05,santoro:06,lidar:08,das:08,morita:08,mukherjee:15}
has sparked a small computing revolution in recent years.  Being a 
novel hardware based on nonsilicon chips used to perform
computations exploiting the potential advantages of quantum fluctuations
\cite{simon:94}, quantum annealing machines might affect the way a
multitude of hard optimization problems are solved today.

The first somewhat useful programmable commercial devices that attempt
to exploit this unique power are the D-Wave One and Two quantum
annealers \cite{johnson:11}, that are designed to solve quadratic
unconstrained binary optimization (QUBO) problems \cite{lucas:14}, such
as finding the ground state of a disordered Ising spin-glass
Hamiltonian, a well-known NP-hard problem in this general formulation
\cite{barahona:82}.  Because many problems across disciplines can be
mapped onto QUBOs, multiple studies of the D-Wave quantum annealer's
performance, compared to some classical optimization approaches, such as
simulated annealing (SA) \cite{kirkpatrick:83}, have been performed
\cite{boixo:13a,boixo:14,wang:13,pudenz:13,vinci:14,shin:14,ronnow:14a,smolin:14,smith:13,venturelli:15a,albash:15,albash:15a}.
Tests \cite{boixo:14,wang:13,vinci:14,smolin:14} by different research
teams suggest that the D-Wave quantum annealer does benefit from quantum
effects. However, it is unclear if this quantum advantage is involved in
the optimization of cost functions.  Furthermore, to date these studies
reveal no clear evidence of limited quantum speedup \cite{ronnow:14a}
over classical optimization algorithms on traditional computers.

Recent work by Katzgraber {\em et al.}~\cite{katzgraber:14} suggests
that current benchmarking approaches using spin glasses with
uniformly-distributed disorder on the Chimera graph \cite{bunyk:14},
such as bimodal or range-$k$, might not be the best benchmark problems
in the quest for quantum speedup.  In particular,
Ref.~\cite{katzgraber:15} proposes an innovative approach based on
insights from the study of spin glasses to design hard benchmark
problems {\em within} the constraints of the D-Wave device.  To overcome
the limitations posed by the D-Wave architecture,
Ref.~\cite{katzgraber:15} proposes to use instances with a unique ground
state, as well as many metastable states.  In this work we study the
interplay between the generation of hard benchmark instances with the
design of problems suitable for the D-Wave device that are robust to
noise. Ideally, thus, a two-tier (unfortunately
computationally-expensive) data mining approach is needed to produce
ideal test instances for any quantum annealing device: First, random
benchmark instances are mined for their desired properties (e.g., unique
ground state) that make them hard problems to solve. Second, these
instances are tested for their robustness to the intrinsic noise present
in any hardware device.

The fact that different numerical studies
\cite{das:05,santoro:06,morita:08,nagaj:12} demonstrated that QA might
outperform SA in certain problems---especially those with rough energy
landscapes---has motivated the authors of Refs.~\cite{katzgraber:15} and
\cite{hen:15a} to design tunable hard benchmarking problems. Reference
\cite{katzgraber:15} goes a step further, by being able to carefully
tune the barrier thickness between dominant features in the energy
landscape, thus putatively allowing for the detection of any quantum
advantage that a quantum annealing device might pose over traditional
optimization approaches.  Despite these efforts, noise due to thermal
excitations and control errors on qubits and couplers have a detrimental
effect on the performance of the D-Wave quantum annealer
\cite{dickson:13,lanting:14,pudenz:13,pudenz:15,correll:15} that likely
is masking any potential limited quantum speedup \cite{ronnow:14a}. A
simple explanation for these problems is given by the fragility of spin
glasses to small perturbations, also known as chaotic effects
\cite{mckay:82,bray:87,kondor:89,ritort:94,neynifle:97,neynifle:98,billoire:00,billoire:02,sasaki:05,katzgraber:07}
to either couplers (bond chaos), qubits via longitudinal fields (field
chaos), or both couplers and qubits (temperature chaos).  Here, small
fluctuations can produce large changes in the free energy of the system,
thus perturbing the original problem Hamiltonian to be solved.

Although quantum error correction \cite{pudenz:13,pudenz:15,correll:15}
can, in principle, mitigate these errors, it does so at a cost of
needing multiple physical qubits to encode one logical qubit, thus
reducing the effective system size of problems to be studied. This also
means that ``error-corrected'' benchmark instances, while more robust to
noise, will likely be too small to be in the scaling regime of interest
for currently available system sizes.  As such, designing hard benchmark
instances that are robust to noise and require no overhead in the
embedding to keep the problem size at a maximum are of utmost importance
to detect quantum speedup. In this work we classically study {\em
resilience}, i.e., the probability that the ground-state configuration
is not affected by random fields and random-bond fluctuations found on
the chip for different benchmark instance classes, by using realistic
uncorrelated noise models for the D-Wave Two quantum annealer.
Furthermore, we present strategies on how to develop hard benchmark
instances that, at the same time, are robust to noise. Note that our
methodology is generic, i.e., it can be applied to any architecture or
noisy black-box optimization device. Furthermore, the study can be
trivially extended to include low-lying excited states if the gold
standard of finding the exact ground state is relaxed to include a
subset of low-lying excited states.

The paper is structured as follows. In Sec.~\ref{sec:model} we introduce
the different benchmark instance classes studied, as well as the noise
model. Furthermore, we describe the heuristic used to find the
ground-state configurations. Our numerical results on the D-Wave chimera
topology are presented in Sec.~\ref{sec:res}, followed by concluding
remarks.

\section{Model, Observables, Algorithm}
\label{sec:model}

Our calculations are for the currently-available D-Wave Two device
\cite{comment:d-wave}. However, the ideas can be generalized to any
topology.

\subsection{Model}

\begin{figure}[tb]
\includegraphics[width=\columnwidth]{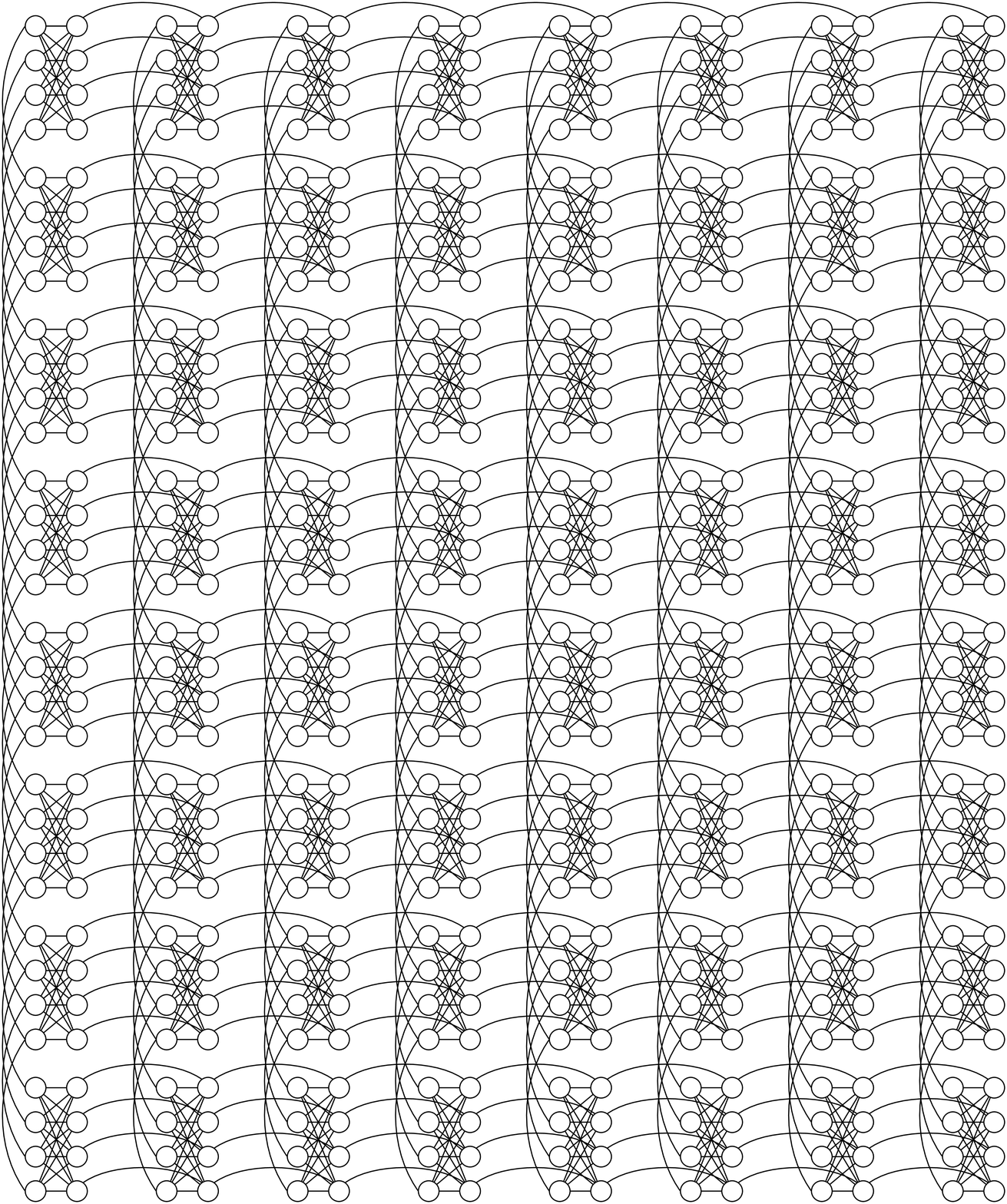}
\caption{
Adjacency matrix of the D-Wave Two chip with $8 \times 8$ K$_{4,4}$
cells and $512$ qubits (circles) connected by couplers (lines).
}
\label{fig:chimera}
\end{figure}

The {\em native} benchmark for the D-Wave Two quantum annealer is an
Ising spin glass \cite{binder:86,nishimori:01,stein:13} defined on the
Chimera topology of the system \cite{bunyk:14}. The Hamiltonian of the
problem to be optimized is given by
\begin{equation}
{\mathcal H} =  -\sum_{\{i,j\} \in {\mathcal V}}J_{ij}s_i s_j 
                - \sum_{i \in {\mathcal V}} s_i h_i \, .
\label{eq:ham}
\end{equation}
where $s_i \in\{\pm 1\}$ signify Ising spins on the vertices ${\mathcal
V}$ of the Chimera lattice. Figure \ref{fig:chimera} shows a $512$ qubit
Chimera lattice with $8 \times 8$ K$_{4,4}$ cells. In addition, each
spin $s_i$ is coupled to a local random field $h_i$. The sum is over all
edges ${\mathcal E}$ connecting vertices $\{i,j\} \in {\mathcal V}$..
The interactions $J_{ij}$ between the spins are drawn from carefully
chosen, discrete disorder distributions within the hardware constraints
of the D-Wave Two architecture.

To emulate the effects of thermal noise in the device, we perturb the
discrete values of the couplers $J_{ij}$ by a random amount $\Delta
J_{ij}$ drawn from a Gaussian distribution with zero mean and standard
deviation $\Delta J$. For simplicity, we assume the noise is quenched
and uncorrelated. This ``white noise'' represents a realistic
(classical) noise model for coupler fluctuations that is typically used
to study the effects of noise in electronic devices, as well as
telecommunications. Although the qubit noise in the D-Wave Two device is
closer to $1/f$ noise with a ``pink'' power spectrum, for simplicity we
couple the individual qubits to uncorrelated quenched random fields
drawn from a Gaussian distribution with zero mean and standard deviation
$h$. We do not expect this simplification to qualitatively change our
results.

\subsection{Instance classes \& observables}

Carefully-chosen interactions between the spins determine the hardness
and robustness of instance classes \cite{katzgraber:15}.  To develop
hard instances, multiple requirements have to be fulfilled.  First, it
is of paramount importance to ensure that the instances have a {\em
unique} ground-state configuration that minimizes the cost function in
Eq.~\eqref{eq:ham}. Furthermore, it is desirable to have dominant
metastable states such that the system is easily trapped -- a process
that can be accomplished by a post-processing selection and mining of
the data based on insights from the study of the dynamics of spin
glasses using classical simulation techniques
\cite{yucesoy:13,katzgraber:15}. Ultimately, an ideal benchmark instance
is robust to noise, has a unique ground state and, ideally, many
metastable states.

To gauge the fraction of unique ground-state configurations for a
particular instance class, we define a quantity we call {\em yield}
($\mathcal Y$), i.e.,
\begin{equation}
{\mathcal Y}=N_{\rm{unique}}/N_{\rm{total}} .
\label{eq:yield}
\end{equation}
In Eq.~\eqref{eq:yield} $N_{\rm{total}}$ is the total number of {\em
randomly-generated} instances for that particular instance class and
$N_{\rm{unique}}$ is the number of instances featuring a unique ground
state (no degeneracy).

One simple approach pioneered in Ref.~\cite{katzgraber:15} to design
instance classes with high yield, is to ensure that as few qubits $s_i$
as possible have zero local fields ${\mathcal F}_i = \sum_{j\neq i}
J_{ij}s_j + h_i$. If for a given qubit ${\mathcal F}_i \equiv 0$, then
the qubit's value does not change the energy of the system. Therefore,
if a system with $N$ qubits has $k$ free qubits with zero local field,
the degeneracy of the ground state is increased by a factor $2^k$.  We
have exhaustively computed the probability that a particular combination
of two, three, or four integer values \cite{comment:4} in the range
$\{\pm 1,\ldots, \pm i_{\rm max}\}$ (with $i_{\rm max} = 28$)
\cite{comment:28} for the couplers $J_{ij}$ on the Chimera topology
yields the smallest fraction of qubits with zero local fields.
Furthermore, we have attempted to ``spread out'' the integers as much as
possible in the range $[-1,1]$ after a normalization of the coupler
values with $i_{\rm max}$. In addition to the previously-studied cases
of bimodal disorder, i.e.,
\begin{itemize}
\item[]{U$_1 \in \{\pm 1\}$,}
\end{itemize}
as well as uniform range-$k$ disorder with $k = 4$ \cite{boixo:14,ronnow:14a}
\begin{itemize}
\item[]{U$_4 \in \{\pm 1,\pm 2,\pm 3,\pm 4\}$,}
\end{itemize}
we also study Sidon-type instances \cite{sidon:32,katzgraber:15}, namely
\begin{itemize}
\item[]{U$_{5,6,7} \in \{\pm 5,\pm 6,\pm 7\}$,}
\end{itemize}
which are similar to uniform range-$7$ instances, however only the three
largest integers that form a Sidon set are kept. Finally, we study a larger
Sidon set
\begin{itemize}
\item[]{S$_{28} \in \{\pm 8,\pm 13,\pm 19,\pm 28\}$.}
\end{itemize}
The U$_{5,6,7}$ and S$_{28}$ Sidon instance classes reduce the
probability of zero local fields drastically by design, and thus
maximize the yield of unique ground states. In fact, while U$_1$ has an
average probability of $23$\% to have zero local fields, this number is
reduced to $6$\% in the U$_4$ class. U$_{5,6,7}$ has only $4.5$\% zero
local fields and S$_{28}$ has $1.5$\%. 

To increase the resilience to noise for a given instance, one has to
maximize the change in energy when flipping a spin, i.e., the minimum
classical energy gap. Ideally, this change in energy should be
considerably larger than the typical noise fluctuations to prevent qubit
errors.  For Ising spins, this energy gap is given by $\Delta E =
2/i_{\rm max}$, where $i_{\rm max}$ is the largest integer in the
unnormalized bond distribution. For example, $\Delta E({\rm U}_1) = 2$,
whereas $\Delta E({\rm U}_4) = 1/2$, $\Delta E({\rm U}_{5,6,7}) = 2/7$,
and $\Delta E({\rm S}_{28}) = 1/14 \sim 0.07$. For the current D-Wave
Two machine with $512$ qubits, coupler fluctuations are typically $\sim
0.035$ if the bonds are normalized to unity (``autoscaling mode'').
This means that in this case the S$_{28}$ instance class pushes the
limits of the machine because $\Delta E({\rm S}_{28}) \sim 2 \Delta J$.

To quantify the robustness of ground-state configurations to noise, we
define the resilience $R$ of an instance to be
\begin{equation}
R = N_{\rm{same}}/N_{\rm{trials}}
\label{:resilience}
\end{equation}
where $N_{\rm{same}}$ is the number of trials with different random
noise perturbations (either fields or bonds) that do not change the
original ground-state configurations. We perform $N_{\rm{trials}}=10$
trials (or gauges) to compute $R$. The resilience of an instance class
is the resilience for each instance $R$ averaged over the bond disorder,
i.e., $\mathcal{R} = [R]_{\rm av}$, where $[\cdots]_{\rm av}$ represents
an average over multiple random bond configurations. A preference should
be given to whole instance classes with high resilience. However,
individual instances that are unaffected by the perturbations are also
robust instances and can be used for benchmarking purposes. Conversely,
to study the effects of noise in quantum annealing machines and how to
reduce these, instances with a {\em small} resilience can also be mined
\cite{venturelli:15a}. 

Finally, we emphasize that a ``relaxed'' resilience $R_k$ can also be
defined, where
\begin{equation}
R_k = N_{\rm{same}}(E \le E_k)/N_{\rm{trials}} .
\label{:resiliencerel}
\end{equation}
Here $N_{\rm{same}}(E \le E_k)$ is the number of times a state with an
energy $E$ less or equal than the energy of the $k$-th excited state is
found. This is of importance when the analog machine suffers from high
noise levels and where the determination of the exact ground state is
difficult or even impossible. For discrete disorder distributions---as
commonly used on quantum annealing machines with finite precision---the
energy levels are separated by well-defined values, i.e., computing the
relaxed resilience of an instance class, $\mathcal{R}_k = [R_k]_{\rm
av}$, is well defined.

\subsection{Algorithm details}

In order to measure the yield and resilience of a particular instance
class, ground states of instances from all instance classes have to be
found. We apply a heuristic method that uses the parallel tempering
Monte Carlo algorithm \cite{hukushima:96} combined with isoenergetic
cluster moves \cite{zhu:15b} to speed  up the thermalization.  Simulation
parameters are listed in Table \ref{tab:params_1} and thermalization has
been determined by a logarithmic binning of the data. Once the last
three bins agree within error bars, we deem the system to be in thermal
equilibrium.  The detailed algorithm to detect ground states was first
introduced in Ref.~\cite{katzgraber:03}. However, to increase the
accuracy of our heuristic, here four instead of two copies of the system
with the same disorder are simulated with {\em independent} Markov
chains.  We perform $N_{\rm{sw}}$ updates \cite{comment:update}.  For
$N_{\rm{sw}}/8$ updates we keep track of the lowest energy $E$ of each
Markov chain at the lowest temperature simulated.  If $E^{(1)} = E^{(2)}
= E^{(3)} = E^{(4)}$, it is very likely the ground state energy $E_0$
has been found.  For the remaining number of updates we keep statistics
of the configurations that minimize the Hamiltonian and thus estimate
the degeneracy distribution of the ground state.  However, there is no
guarantee that any solution obtained by this heuristic method is the
true optimum, or that we have found all configurations that minimize the
Hamiltonian. Fortunately, for the Sidon-type instance classes the
degeneracy is small by construction. Therefore, it is likely that we
found all ground-state configurations.  Once the ground-state
configurations of all instances have been found, the average yields for
different instance classes can be computed.

In addition to the effects of the minimum energy gap $\Delta E$ on the
resilience for each instance class, we also consider the effects of the
number of first excited states on the resilience. To estimate the number
of first excited states, for the remaining $(7/8)N_{\rm{sw}}$ sampling
updates we also keep track of all configurations that have an energy
$E_1 = E_0 + \Delta E$.

\begin{table}
\caption{
Simulation parameters: For each instance class and system size $N$, we
compute $N_{\rm sa}$ instances. $N_{\rm sw} = 2^b$ is the total number
of Monte Carlo sweeps for each of the $4 N_T$ replicas for a single
instance, $T_{\rm min}$ [$T_{\rm max}$] is the lowest [highest]
temperature simulated, and $N_T$ is the number of temperatures used in
the parallel tempering method. For the lowest $N_{\rm icm}$ temperatures
isoenergetic cluster moves are applied.
\label{tab:params_1}
}
\begin{tabular*}{\columnwidth}{@{\extracolsep{\fill}} l l l l l l l r }
\hline
\hline
Class  & $N$ & $N_{\rm sa}$ & $b$ & $T_{\rm min}$ & $T_{\rm max}$ & $N_{T}$ &$N_{\rm icm}$  \\
\hline
U$_{1}$		& $512$  & $900$ & $19$ & $0.150$ & $3.050$ & $30$ &$13$ \\
\hline
U$_{4}$		& $512$  & $900$ & $19$ & $0.150$ & $3.000$ & $30$ &$14$\\
\hline
U$_{5,6,7}$	& $128$  & $900$ & $19$ & $0.150$ & $3.000$ & $30$ &$14$\\
U$_{5,6,7}$	& $288$  & $900$ & $19$ & $0.150$ & $3.000$ & $30$ &$14$\\
U$_{5,6,7}$	& $512$  & $900$ & $19$ & $0.150$ & $3.000$ & $30$ &$14$\\
U$_{5,6,7}$	& $800$  & $900$ & $19$ & $0.150$ & $3.000$ & $30$ &$14$\\
U$_{5,6,7}$	& $1152$ & $900$ & $19$ & $0.150$ & $3.000$ & $30$ &$14$\\
\hline
S$_{28}$	& $512$  & $900$ & $19$ & $0.150$ & $3.000$ & $30$ &$14$ \\
\hline
\end{tabular*}
\end{table}

\section{Results}
\label{sec:res}

We focus only on the resilience of the exact ground state in this study
for the sake of brevity and to illustrate the developed methodology. Our
approach is easily extended to include low excited states. Note that if
the resilience $R$ of an instance is large, we also expect the relaxed
instance resilience $R_k$ to be large for small enough $k$.

\subsection{Yield of non-degenerate ground states}

For the current D-Wave Two architecture with $512$ qubits, the yield of
unique ground states is strongly dependent on the instance class
(disorder between spins) used.  When the disorder is drawn from a
bimodal distribution (U$_1$) the yield in all our experiments was
exactly $0$\%. Surprisingly, uniform range-$4$ instances (U$_4$) also
have ${\mathcal Y} = 0$\%. However, by increasing the range of the
integers and selecting them from a Sidon set while removing the lowest
values gives ${\mathcal Y} = 4.5(4)$\% for the U$_{5,6,7}$ class.
Although a small fraction, it is clearly nonzero. Finally, for the large
Sidon set S$_{28}$ we obtain a fraction ${\mathcal Y} = 20.0(6)$\% of
unique ground states \cite{comment:error}, i.e., optimal for
large-scale benchmarking.

\subsection{Resilience to noise}

\begin{figure}[tb]
\includegraphics[width=\columnwidth]{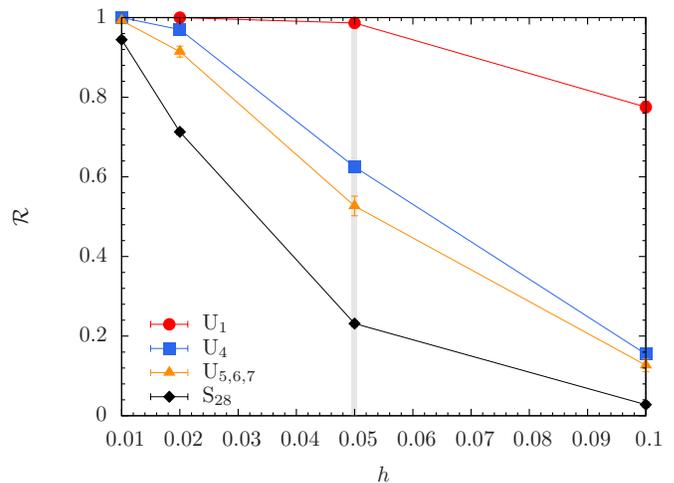}
\caption{(Color online)
Resilience (${\mathcal R}$) of different instance classes (see text) for
a $N=512$ qubit system on the Chimera graph as a function of Gaussian
random field strength ($h$).  Instance classes are less resilient to
noise with increasing field strength and decreasing classical energy
gap. The shaded line represents the current field noise strength of
approximately $5$\% in the D-Wave Two system.
}
\label{fig:field_512}
\end{figure}

Figure \ref{fig:field_512} shows the resilience to random-field noise
for different instance classes. As the typical field strength $h$
increases, the resilience ${\mathcal R}$ for all instance classes
decreases.  This is to be expected, because the energy spread due to the
splitting of degenerate excited states via the random fields results in
more energy levels crossing.  Furthermore, for a fixed field strength,
instance classes with small energy gaps $\Delta E$ tend to have lower
resilience.  This is to be expected: it is easier for split states to
have a lower energy than the original ground state when the gap is
small.  Note that while instance classes U$_{5,6,7}$ and U$_4$ have a
similar resilience, the yield of unique ground states is considerably
higher for U$_{5,6,7}$, i.e., a careful design of the spin-spin
interactions is key when attempting to benchmark a quantum annealing
device.

\begin{figure}[tb]
\includegraphics[width=\columnwidth]{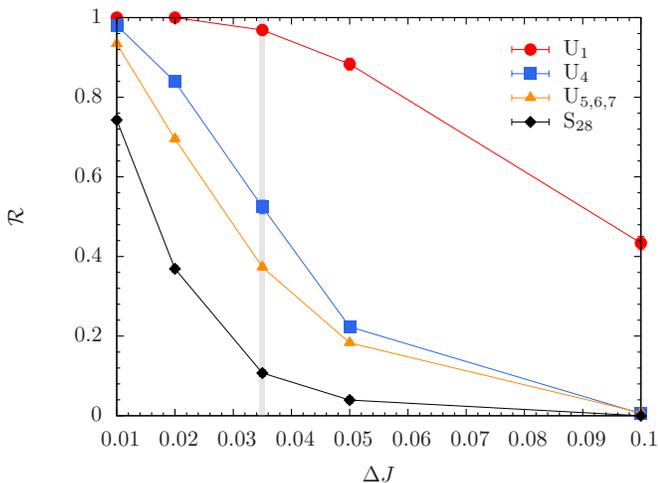}
\caption{(Color online)
Resilience (${\mathcal R}$) of different instance classes (see text) for
a $N=512$ qubit system on the Chimera graph as a function Gaussian
random bond fluctuation strength ($\Delta J$).  Instance classes are
less resilient to noise with increasing bond fluctuation strength and a
decreasing classical energy gap. The shaded line represents the current
bond noise strength in the D-Wave Two system, i.e., $\sim 3.5$\%. Note
that bond noise has a stronger effect than field noise
(Fig.\ref{fig:field_512}) on the device.
}
\label{fig:bond_512}
\end{figure}

Figure \ref{fig:bond_512} shows the resilience of different instance
classes as a function of different typical coupler perturbations $\Delta
J$. Again, for all instance classes studied, the resilience decreases as
fluctuations increase. In addition, instance classes with small energy
gaps have a lower resilience. It is important to note that bond noise
has a stronger impact on the resilience than field noise. Considering
each qubit has typically $\sim 6$ neighbors in the Chimera lattice, the
impact of bond noise is amplified by multiple connections of qubits.
Therefore, reducing the fluctuations of the couplers is more important
than dealing with the intrinsic flux noise of each qubit.

Unfortunately, for the D-Wave architecture, to find an instance class
that is both hard and robust to noise, compromise has to be made.  The
U$_1$ instance class has the highest resilience to noise, however, the
huge ground-state degeneracy makes it easier for classical algorithms
such as SA to find minimum-energy configurations
\cite{katzgraber:14,katzgraber:15}.  On the flip side, the Sidon
instance class is known to be hard \cite{katzgraber:15} and produces
many unique ground states, but its resilience is comparably low due to
the small energy gap.  A compromising natural choice would therefore be
to either use the U$_4$ or U$_{5,6,7}$ instance classes.  However, while
the resilience for both U$_4$ and U$_{5,6,7}$ are comparable, the yield
of unique ground states needed to construct hard benchmark problems is
much higher for U$_{5,6,7}$. We thus conclude that for the current
Chimera topology, the U$_{5,6,7}$ instance class is the optimal
compromise to design hard benchmark problems within the D-Wave Two
architecture constraints. For the remainder of this paper we thus focus
on this particular instance class.

\begin{figure}[tb]
\includegraphics[width=\columnwidth]{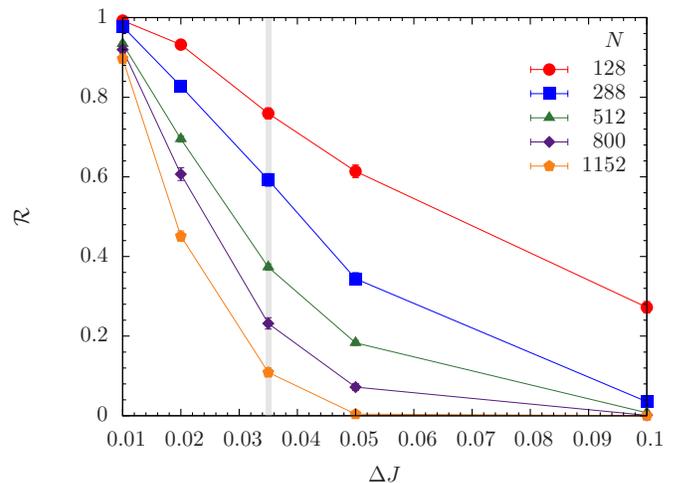}
\caption{(Color online)
Resilience ${\mathcal R}$ of the U$_{5,6,7}$ instance class as a
function of the bond fluctuation strength ($\Delta J$) for different
system sizes $N$ on the Chimera topology.  The resilience clearly
decreases for increasing noise and system size.  The shaded vertical line
represents the current bond-noise strength in the D-Wave Two system,
approximately $3.5$\%.
}
\label{fig:567}
\end{figure}

Figure \ref{fig:567} shows the resilience of the U$_{5,6,7}$ instance
class for different system sizes $N$ of the Chimera lattice as a
function of the random-bond fluctuation strength $\Delta J$.  Clearly,
for increasing system size the resilience ${\mathcal R}$ decreases
(larger system sizes typically have a higher degeneracy, therefore
level crossings are more common than with smaller systems).  This
means that to scale up the system size of the D-Wave Two---or any
other quantum annealing device---in the future, a much more precise
control over the device's noise and/or the implementation of error
correction schemes \cite{pudenz:13,pudenz:15,correll:15} are
imperative.

We conclude this section by quoting results specifically calculated for
the current D-Wave Two ($512$ theoretical qubits) and the
next-generation D-Wave 2X ($1152$ theoretical qubits) machines when both
errors in the couplers and qubits are applied, using the real values
provided by D-Wave, Inc.~\cite{comment:aqc}.  For the D-Wave Two machine
with $512$ qubits, $\Delta J = 3.5$\% and $\Delta h = 5$\%. Applying
both coupler and qubit perturbations yields an average resilience of
$\mathcal{R} = 0.22(2)$.  For the next-generation D-Wave 2X device noise
levels have been reduced, i.e., $\Delta J = 2.5$\% and $\Delta h = 3$\%.
This results in $\mathcal{R} = 0.21(3)$. We point out two interesting
facts: First, it seems that the resilience for both coupler and qubit
noise is approximately the product of the resilience of only noise being
considered on the couplers with the resilience of only noise being
considered on the qubits. Thus, as a rule of thumb and to obtain an
approximate estimate for the combined effects, the individual numbers
can be multiplied. Second, despite the lower noise level of the
next-generation device, the resilience remains approximately unchanged
within error bars. It seems that the increased number of qubits cancels
out the additional precision.

\subsection{Effects of the number of first excited states}

\begin{figure}[tb]
\includegraphics[width=\columnwidth]{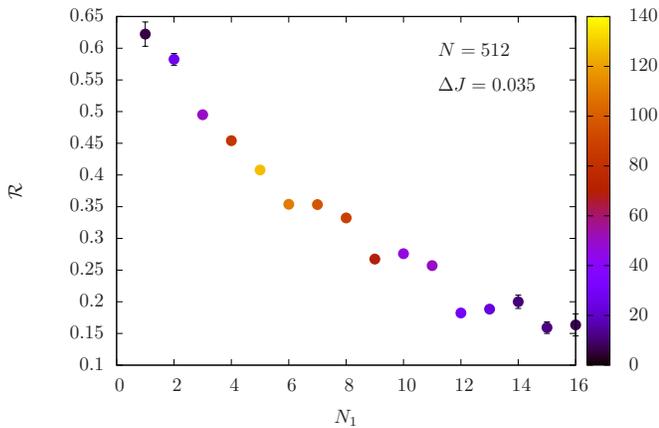}
\caption{(Color online)
Resilience ${\mathcal R}$ as a function of the number of first excited
states $N_1$ for $N=512$ spins on the Chimera lattice. The data are for
the U$_{5,6,7}$ instance class. The color bar shows approximately how
often a given number of first excited states occurs for the $900$
instances studied. In this case, between four and eight first excited
states are most common.
}
\label{fig:N1}
\end{figure}

Figure \ref{fig:N1} shows the resilience ${\mathcal R}$ of the
U$_{5,6,7}$ instance class as a function of the degeneracy of the first
excited state on the Chimera topology with $N = 512$ spins.  The higher
the degeneracy of the first excited state, the lower the resilience.
This can be explained by the increased probability of level crossing.
We also color coded each dot in the figure: The heat map represents the
number of instances that had a given degeneracy $N_1$ of the first
excited state out of the $900$ simulated. In this case, the bulk of the
instances have between $4$ and $8$ degenerate first excited states.
This results in a reduction of the resilience, compared to instances
that contain only one or two first excited states.

While instances with only one or two first excited states are extremely
rare, the effort needed to find these might outweigh the approximately
$30$\% in the resilience reduction by allowing states with three to four
first excited states. We thus recommend to fix the number of first
excited states to be less than or equal to four in this case.

We have also computed the Hamming distance between the ground state and
all first excited states for a given instance. Our results suggest that
when the average Hamming distance is small, the resilience to noise is
higher. A simple explanation is that both ground-state and excited
configurations are quite similar and therefore the noise affects them
comparably, i.e., both the ground state and the first excited states are
shifted approximately by the same amount when the Hamming distance is
small.

\section{Conclusions}
\label{sec:conclusions}

In order to develop both hard and robust benchmark instances, we have
tested different instance classes by computing their yield (fraction of
instances with a unique ground-state configuration) and resilience to
noise fluctuations.  Ideally, hard instances (high yield) with a high
resilience are optimal for benchmarking purposes. Both yield and
resilience can be tuned by a careful design of the instance
classes---within the hardware restrictions of the machine---followed by
a mining of the data. Although the numerical effort to do such
``designer instances'' is nonnegligible, we think this is a key
ingredient in designing good benchmarks for quantum annealing devices,
as well as any other computing architectures. It seems that both
resilience and yield for the Chimera topology are slightly
anticorrelated. A good compromise is thus the U$_{5,6,7}$ instance class
where $J_{i,j} \in \{\pm 5, \pm 6, \pm 7\}$ that has a good resilience
to both field and coupler noise, as well as a nonzero yield of unique
ground states, with a small number of first excited states.

We emphasize that our results for the resilience represent a {\em
best-case scenario} for any quantum annealing machine. Any other source
of error can only decrease the success probabilities further.  However,
it could be that the introduction of carefully-crafted correlations
between bond and field noise might reduce the errors and increase the
resilience.  Bond noise is the most limiting issue for the current
D-Wave Two quantum annealer and is highly dependent on the connectivity
of the graph. While it is desirable to have a high connectivity to be
able to embed interesting problems on any putative architecture, one has
to also keep in mind that noise levels should be far lower than in the
current D-Wave machine.

This classical study of both resilience and yield plays an important
role in the design of future adjacency matrices for quantum annealing
machines, as well as the study of strategies to reduce noise in quantum
annealers. Our results and methods can easily be generalized to other
systems and thus should be of general interest when designing hard
instance problems that attempt to circumvent the limitations of current
hardware. Furthermore, calibration of future generations of the D-Wave
device should be improved to allow for the encoding of more complex
Sidon sets and thus the design of harder benchmark problems.  Similarly,
although the main goal of this work is to produce problems that are
robust to noise, the methodology presented can be used to design
tailored instances that are particularly sensitive to noise. This could
play an important role when designing approaches to better calibrate
devices, as done in Ref.~\cite{perdomo:15}. Finally, we emphasize that
if either noise is large or the instances produced are too difficult to
minimize, a relaxed resilience that includes low-lying excited states
can be defined.

\begin{acknowledgments} 

We would like to thank M.~H.~Amin, P.~Bunyk,T.~M.~Lanting,
A.~Perdomo-Ortiz and H.~Mu\~{n}oz-Bauza for fruitful discussions. We
also thank H.~Mu\~{n}oz-Bauza for assistance.
H.~G.~K.~acknowledges support from the National Science Foundation
(Grant No.~DMR-1151387) and would like to thank the makers of the
Dish-Wash Dos for motivation on this project. He also would like to
especially thank P.~Bunyk and T.~M.~Lanting for explaining multiple
details related to the D-Wave Two device.  H.~G.~K.~and S.~S.~would like
to thank the European Commission through the IRSES network DIONICOS
under Contract No.~PIRSES-GA-2013-612707 (FP7-PEOPLE-2013-IRSES).  We
would like to thank the Texas Advanced Computing Center (TACC) at The
University of Texas at Austin for providing HPC resources (Lonestar
Linux Cluster) and Texas A\&M University for access to their Eos and Ada
clusters.  This research is based upon work supported in part by the
Office of the Director of National Intelligence (ODNI), Intelligence
Advanced Research Projects Activity (IARPA), via MIT Lincoln Laboratory
Air Force Contract No.~FA8721-05-C-0002.  The views and conclusions
contained herein are those of the authors and should not be interpreted
as necessarily representing the official policies or endorsements,
either expressed or implied, of ODNI, IARPA, or the U.S.~Government.
The U.S.~Government is authorized to reproduce and distribute reprints
for Governmental purpose.

\end{acknowledgments}

\bibliography{refs,comments}

\begin{thebibliography}{64}
\bibitem[{com({\natexlab{b}})}]{comment:4}
\bibinfo{note}{We restricted the search to four integer values because of the
  hardware restrictions on the D-Wave Two machine that require the interactions
  to be well separated in the interval $[0,1]$.}
\bibitem[{com({\natexlab{c}})}]{comment:28}
\bibinfo{note}{We chose $i_{\rm max} = 28$ empirically because it allowed us to
  encode very hard problems with many unique ground states.}
\bibitem[{com({\natexlab{e}})}]{comment:error}
\bibinfo{note}{Error bars are computed using a bootstrap analysis.}
\end{thebibliography}

\end{document}